\documentclass{article}

\usepackage{arxiv}

\usepackage[utf8]{inputenc} 
\usepackage[T1]{fontenc}    
\usepackage{hyperref}       
\usepackage{url}            
\usepackage{booktabs}       
\usepackage{amsfonts}       
\usepackage{nicefrac}       
\usepackage{microtype}      
\usepackage{graphicx}
\usepackage{natbib}
\usepackage{doi}
\pdfoutput=1

\title{Photon-pair generation in a heterogeneous silicon photonic chip}

\author{ \\Mingwei Jin$^{1,2,\dagger}$,   Neil MacFarlane$^{3,\dagger}$,  Zhaohui Ma$^{1,2}$,  Yongmeng Sua$^{1,2}$\\
Mark A. Foster$^{3}$,  Yuping Huang$^{1,2}$,  Amy C. Foster$^{3}$\\
\\
$\dagger$. Equally contributing authors\\
1. Department of Physics, Stevens Institute of Technology, Hoboken, NJ, 07030, USA\\
2. Center for Quantum Science and Engineering, Stevens Institute of Technology, Hoboken, NJ, USA\\
3. Department of Electrical and Computer Engineering, Johns Hopkins University, 3400 North Charles Street,\\ Baltimore MD, 21218, USA
}

\renewcommand{\headeright}{}
\renewcommand{\undertitle}{}

\hypersetup{
pdftitle={A template for the arxiv style},
pdfsubject={q-bio.NC, q-bio.QM},
pdfauthor={David S.~Hippocampus, Elias D.~Striatum},
pdfkeywords={First keyword, Second keyword, More},
}

\begin{document}
\maketitle

\begin{abstract}
	Integrated Silicon photonics has played an important role in advancing the applications of quantum information and quantum science. However, due to different material properties, it is challenging to integrate all components with excellent performance based on homogeneous material. Here, by combining high nonlinearity and low losses in a heterogeneous silicon platform, we efficiently generate high-quality photon pairs through spontaneous four-wave mixing in hydrogenated amorphous silicon waveguide and route them off-chip through low loss silicon nitride waveguide. A record high coincidence- to- accidental rate value of 1632.6 ($\pm$ 260.4) is achieved in this heterogeneous design with a photon pair generation rate of 1.94 MHz. We also showcase a wide range of multi-channel photon sources with coincidence- to- accidental rate consistently at 200. Lastly, we measure heralded single-photons with a lowest $g^{(2)}_H(0)$ of 0.1085 $\pm$ 0.0014. Our results demonstrate the heterogeneous silicon platform as an ideal platform for efficient generation of photon pairs and routing them off-chip with low losses. It also paves a way for the future hybrid photonic integrated circuit by collecting distinct features from different materials. 
\end{abstract}


\section{Introduction}
Integrated photonics has emerged as a key technology in various applications such as data centers, telecommunication, photonic computing, and quantum computing. These applications require numerous integrated components including low-loss waveguide routing, laser sources, photodetectors, modulators, single photon and photon pair sources. However, due to distinct properties of different materials, it becomes necessary to incorporate various materials in heterogeneous platforms to incorporate different components onto a single platform\cite{Davanco2017,Schnauber2019,Moody_2022}.

Among those photonic components, correlated photon-pair sources are of great interest due to  their applications in quantum computing, quantum teleportation, quantum information processing, quantum sensing, quantum cryptography and others\cite{07quantumcomputing,Bouwmeester1997,Kimble2008,quantumsensing,RevModPhys.74.145,O'Brien2009}. A photon pair source with a broad and switchable bandwidth is essential for applications such as quantum frequency multiplexing and quantum memory\cite{Lavoie2013,spectral}. Over the past decade, there has been tremendous progress with chip-integrated photon pair sources by spontaneous four-wave mixing (SFWM) and spontaneous parametric down-conversion (SPDC), such as those in crystalline silicon (c-Si) \cite{Kues2017,Lu:16}, silicon nitride ($\mathrm{Si_3N_4}$) \cite{Zhang_2016}, hydrogened amorphous Silicon (a-Si:H) \cite{Wang:14,Hemsley2016}, gallium phosphide (GaP) \cite{Logan:18}, gallium arsenide (GaAs) \cite{GaAs}, and lithium niobate (LiNbO3)\cite{zhaohui}. Among these platforms, a-Si:H has achieved particular attention as a photonic platform due to its ultra-high Kerr nonlinearity  ($n_2 = 7.43 \pm 0.87 x 10^{-13}$ $cm^2/W$) comparing to other silicon-based material platforms\cite{asi_kerr}. With such a high Kerr coefficient, efficient SFWM can be achieved with only millimeters of waveguide propagation. Contending with crystalline Silicon (c-Si), a-Si:H can be easily deposited at a lower temperature and thus incorporated in massive chip manufacturing processes\cite{Hemsley2016}.

Despite the excellent Kerr nonlinearity of a-Si:H, the lowest reported a-Si:H linear propagation losses are about an order of magnitude higher than $\mathrm{Si_3N_4}$\cite{asi_loss}. In quantum application, Even though the high nonlinearity of a-Si:H plays a key role in efficiently generating photon pairs, the high propagation loss of a-Si:H waveguide resuts in many of the photon pairs being lost on chip before off-chip detection. This dramatically reduces the coincidence- to- accidental rate that can be observed and severely limits its quantum applications. 

In our previous work, to mitigate higher propagation losses in a-Si:H waveguides, $\mathrm{Si_3N_4}$ waveguides are used for on/off-chip coupling, and vertical coupling between a-Si:H and $\mathrm{Si_3N_4}$ occurs on-chip\cite{multi_layer_result}. With a careful design of vertical coupler, the coupling loss is optimized around 0.5 dB per transition. The replacement of the a-Si:H with a $\mathrm{Si_3N_4}$ waveguide for on-off chip coupling dramatically reduces fiber-to-fiber loss. This reduces the required pump power especially for photon pair generation and broadens the applications of such a platform to other quantum technologies.

In this paper, leveraging the low loss and high nonlinearity of the heterogeneous platform, we further explore the quantum performance of a-Si:H/$\mathrm{SiN_x}$ waveguides. We experimentally demonstrate the photon pair generation with the highest coincidence- to- accidental rate (CAR) of 1632.6 ($\pm$ 260.4) in non-resonant a-Si:H waveguides, showing significant improvement comparing to the previous state of the art value in non-resonant a-Si:H waveguide\cite{Wang:14}. The pure photon pairs are measured at multiple channels with a detuning bandwidth of 6 nm and CAR value consistently at 200. In addition, with the same chip, we perform heralded generation of single photons. The auto-correlation is measurement as low as 0.0126 $\pm$ 0.005. Our work combines the high nonlinearity of a-Si:H and the low propagation loss of $\mathrm{Si_3N_4}$. Our results demonstrate the benefits of a heterogeneous silicon platform for quantum applications by utilizing both low transmission loss and highly nonlinearity. With low-loss inter-layer coupler, such heterogeneous silicon platform also paves the way for integrating even more layers and materials to achieve high-quality components for various applications in quantum applications.


\begin{figure}[htbp]
  \centering
  \includegraphics[width=12cm]{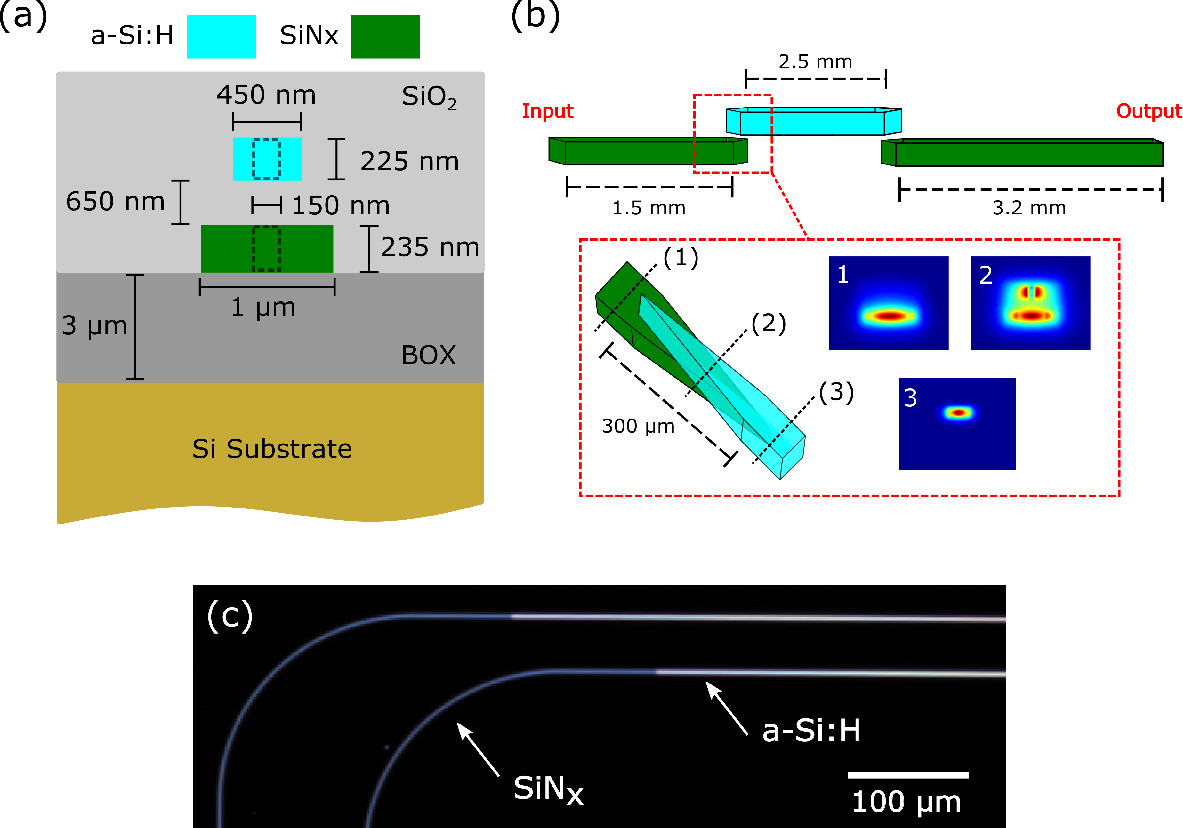}
\caption{\scriptsize{(a) Cross-section of multi-layer platform. (b) Side view of device showing propagation lengths in $\mathrm{\mathrm{SiN_x}}$ (green) and a-Si:H (blue) waveguides. (Inset - Top-down view of interlayer coupler with electric field profiles at three locations along the interlayer coupler). (c) Dark-field microscope image of $\mathrm{\mathrm{SiN_x}}$/a-Si:H waveguides and interlayer coupling region.}}
\label{figure1}
\end{figure}

\section{Experimental Results}


The multi-layer architecture is shown in Figure \ref{figure1} (a-c). The device fabrication starts with a 100-mm silicon wafer with a 3-$\mu$m thick layer of thermal oxide on its surface. 240 nm of $\mathrm{SiN_x}$ is deposited using low-pressure chemical vapor deposition (LPCVD). The recipe parameters consisted of a deposition rate of 4-nm/min, a temperature of 775 $\mathrm{^oC}$, 1000-MPa tensile stress, and 250 mTorr chamber pressure. Platinum marks for alignment between layers were patterned using electron beam lift-off lithography. Weaveguides in the $\mathrm{SiN_x}$ layer were patterned and etched by e-beam lithography and reactive ion etching respectively. The Silicon dioxide ($\mathrm{SiO_2}$) cladding was then deposited with plasma-enhanced chemical vapor deposition (PECVD). Planarization of the $\mathrm{SiO_2}$ surface was performed using chemical mechanical polishing. 225-nm of a-Si:H was then deposited by PECVD. The a-Si:H layer was deposited using recipe parameters of a temperature of 300 $\mathrm{^oC}$, a RF power of 15-W, 1000-SCCM of 5\% silane/helium gas flow, and 1800 mTorr chamber pressure. With the lower a-Si:H deposition temperature we are able to deposit it directly on top of the $\mathrm{SiN_x}$ waveguides without damaging them. Finally, a-Si:H waveguides are patterned and etched using e-beam lithography and reactive ion etching. The a-Si:H waveguides are clad with PECVD ($\mathrm{SiO_2}$) and individual chips are diced out of the wafer. Figure \ref{figure1} (b) shows the geometry of the interlayer coupler. The optical modes at different coupler sections are simulated as figure \ref{figure1} (b). Figure \ref{figure1} (c) shows a dark-field microscope image of $\mathrm{SiN_x}$ and a-Si:H waveguides and the interlayer coupling region.

\begin{figure}[htbp]
  \centering
  \includegraphics[width=12cm]{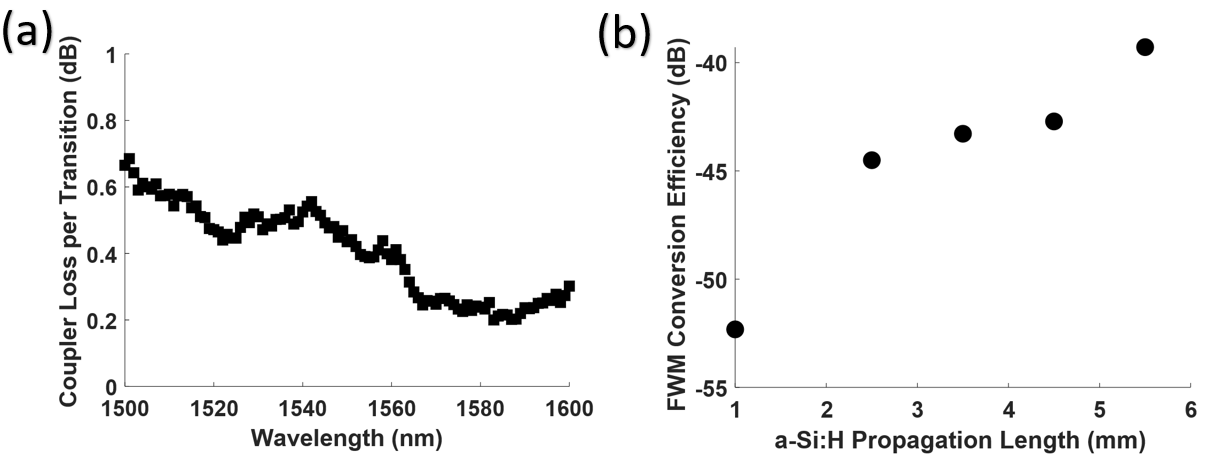}
\caption{\scriptsize{(a) Interlayer coupler loss per wavelength. (b) FWM conversion efficiency for different propagation lengths of a-Si:H waveguides. (originally published in \cite{jhu_stevens_cleo_pgr})}}
\label{figure2}
\end{figure}

To characterize the Kerr coefficient of a-Si:H waveguides, we first perform classic degenerate FWM in our multi-layer devices. Light is coupled on/off chip via inverse taper couplers in the $\mathrm{SiN_x}$ layer as shown in Figure \ref{figure1} (b). We couple on/off the chip using tapered lensed fibers with mode-field diameters of ~2.5 $\mathrm{\mu m}$. The $\mathrm{SiN_x}$ layer offers improved fiber-to-chip coupling efficiency from better mode-matching with the lensed fiber compared to the a-Si:H waveguides. Once on-chip, the light propagates for 1.5 mm in the $\mathrm{SiN_x}$ layer and is then coupled to the a-Si:H layer via evanescent inverse taper couplers and propagates for varying a-Si:H lengths between 1 mm and 5.5 mm. The signal is then coupled back to the $\mathrm{SiN_x}$ layer, where it propagates for another 3.2 mm and is finally output via the $\mathrm{SiN_x}$ layer and coupled to a tapered lensed fiber. The on-chip interlayer coupling insertion loss is measured to be less than 1 dB per transition over a 100 nm bandwidth, while the propagation losses in the $\mathrm{SiN_x}$ and a-Si:H waveguides are 0.5 dB/cm and 4 dB/cm, respectively. The insertion losses as a function of wavelength are shown in Figure \ref{figure2} (a). We measure the four-wave mixing conversion efficiency (CE), defined as the ratio of the output idler power to the input signal power, using multiple devices with varying propagation lengths in the a-Si:H layer. The measured conversion efficiency values as a function of a-Si:H propagation length are shown in Figure \ref{figure2} (b). These results equate to a Kerr coefficient of $\sim 5$ x $10^{-13}$ $\mathrm{cm^2/W}$, which is comparable to the state-of-the-art value ($7.43$ x $10^{-13}$ $\mathrm{cm^2/W}$) \cite{asi_kerr}.


\begin{figure}[htbp]
  \centering
  \includegraphics[width=12 cm]{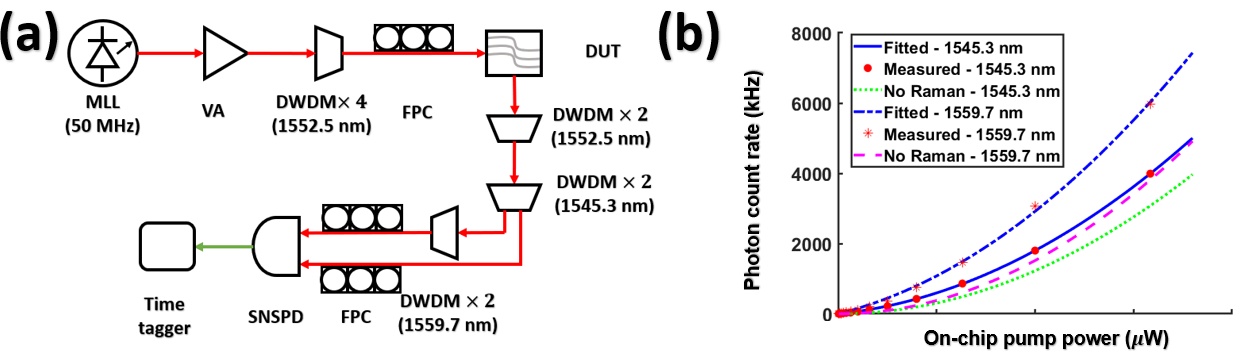}
\caption{(a) Experimental setup. MLL: mode locked laser. VA: variable attenuator. DWDM: dense wavelength-division multiplexer. FPC: fiber polarization controller. DUT: device under test. SNSPDs: superconducting nanowire single photon detectors. (b) Single photon counts of Stokes and anti-Stokes channels versus the on-chip pump power. ((b) originally published in \cite{jhu_stevens_cleo_pgr})}
\label{setup}
\end{figure}

The photon pairs are then generated through SFWM in the a-Si:H waveguides. 
In figure \ref{setup} (a), a 50MHz mode lock laser(Calmar Mendocino) is connected with a variable attenuator
and four cascaded dense wavelength-division multiplexing (DWDM) filters at 1552.5 nm to accurately control pump power and suppress photons beyond pump wavelength, respectively.
We couple to the TE waveguide mode for optimum SFWM efficiency and broad bandwidth by carefully tuning a fiber polarization controller (FPC) and couple to the $\mathrm{SiN_x}$ inverse taper coupler through a lensed fiber.
After propagating in the $\mathrm{\mathrm{SiN_x}}$ waveguide, light is coupled through an interlayer coupler to a 2.5 mm long a-Si:H waveguide where the photon-pair  is generated. The generated photon-pair is coupled back to a low-loss $\mathrm{SiN_x}$ waveguide and collected by a lensed fiber at the output. The total on-off chip loss is 9 dB.
The remaining pump photons are immediately rejected by two cascaded DWDM filters with labelled 60 dB extinction.
Two DWDM filters at 1545.3 nm collect Stokes photons and reject all other wavelengths. The anti-Stokes photon at 1559.7nm is collected by two other DWDMs. The bandwidth for the DWDM filters are aligned to be 200 GHz. The total losses by the output DWDM sets are 1.9 dB and 2.2 dB at Stokes and anti-Stokes channels, respectively. To ensure maximum detection efficiency, two FPCs are placed before entering superconducting nanowire single photon detectors (4 channels SNSPDs, ID281, ID Quantique).    
The SNSPDs have dark-count rate of 50-100 Hz and detection efficiency of 85\%.
The detected signals are lastly resolved by a time-tagging unit (Swabian Instruments) for photon counting and correlated detection.

The stimulated Raman scattering (SRS) noise is first characterized by measuring the power-dependent single photon counts. In the waveguide, the SRS generated photons are linearly proportional to the on-chip pump power $P_{in}$ while the SFWM generated photons are proportional to $P_{in}^2$. Thus, SRS noise can be abstracted at Stokes and anti-Stokes channels by fitting the curves as shown in figure \ref{setup} (b).

\begin{figure}[htbp]
  \centering
  \includegraphics[width=12cm]{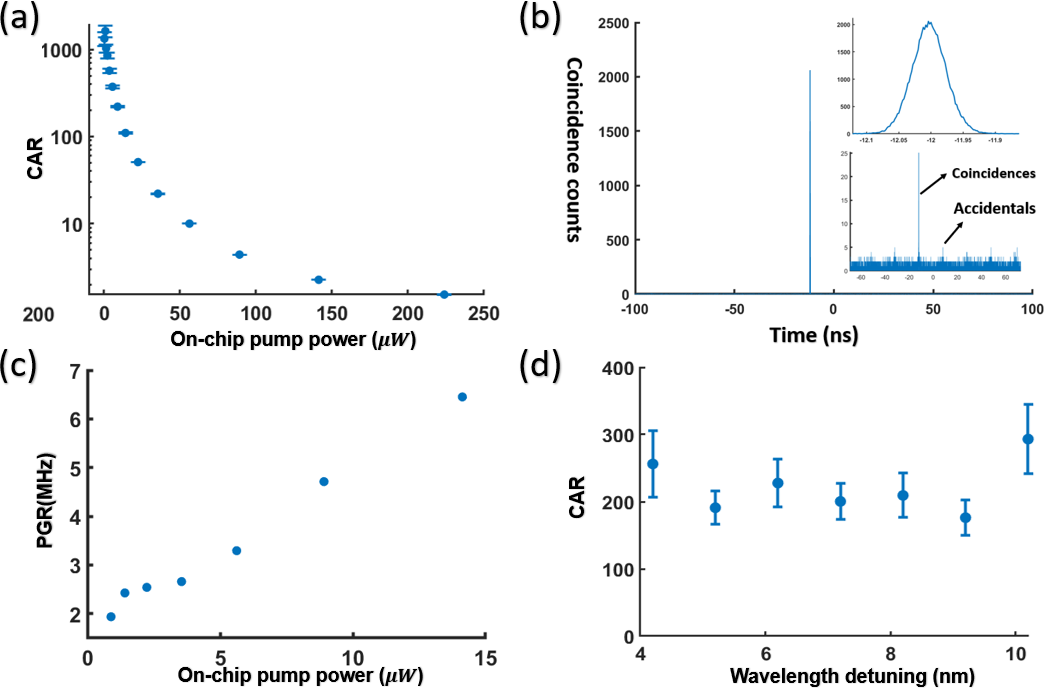}
\caption{\scriptsize{ (a) CAR versus the on-chip pump power. (b) Coincidences measurement at CAR of  1632.6 ($\pm$ 260.4). The upper inset shows the coincidences at zero time delay. The lower inset shows the background noise and accidental at non-zero time delays. (c) Photon pair generation rate (PGR) at low power regime. (d) CAR for different wavelength channels. ((a) and (d) originally published in \cite{jhu_stevens_cleo_pgr})}}
\label{figure3}
\end{figure}

Figure \ref{figure3} (a) shows the coincidence-to-accidental ratios (CAR) versus on-chip pump power. The dark count has been subtracted from coincidence and accidental counts to eliminate dark-count noise, which may significantly decrease CAR in the low count rate regime. The highest CAR is achieved at 1632.6 ($\pm$ 260.4) with average on-chip pump power of about -30.5 dBm. The corresponding photon pair generation rate (PGR) is measured and estimated to be 1.94 MHz. The CAR then drops to 1335.3 ($\pm$ 240.9) when further lowering on-chip pump power. This occurs because the background noise contributes more to the result when the count rates of Stokes and anti-Stokes photons are closer to the dark-count level. Figure \ref{figure3} (b) plots coincidence measurement at the highest measured CAR. The coincidences are counted using the full width at half maximum (FWHM) of the pulse which is depicted in the upper inset. The lower inset shows the accidental counts and background noise. Figure \ref{figure3}(c) plots PGR at low on-chip pump power regime. The error bars are within the dotted areas. The coincidence rate is measured on the same run of CAR measurement. The PGR is then estimated as $N_1N_2/N_{12}$, where $N_1$, $N_2$ and $N_{12}$ donates the detection rate of signal, idler and two-photon coincidences.

Figure \ref{figure3} (d) depicts CAR measurement for multi-channel photon-pairs. To measure the CAR at different wavelength channels, instead of using DWDMs to select the signals and idlers, the photon pairs are filtered using a programmable waveshaper. The waveshaper introduces an additional 5 dB loss, which increases the required pump power. With on-chip pump power at -20.5 dBm, the CAR remains about 200 for the wavelength detuning from 4.2 nm to 10.2 nm. This showcases a wide bandwidth for photon-pair generation that can be dispersion engineered into these devices.

\begin{figure}[htbp]
  \centering
  \includegraphics[width=12cm]{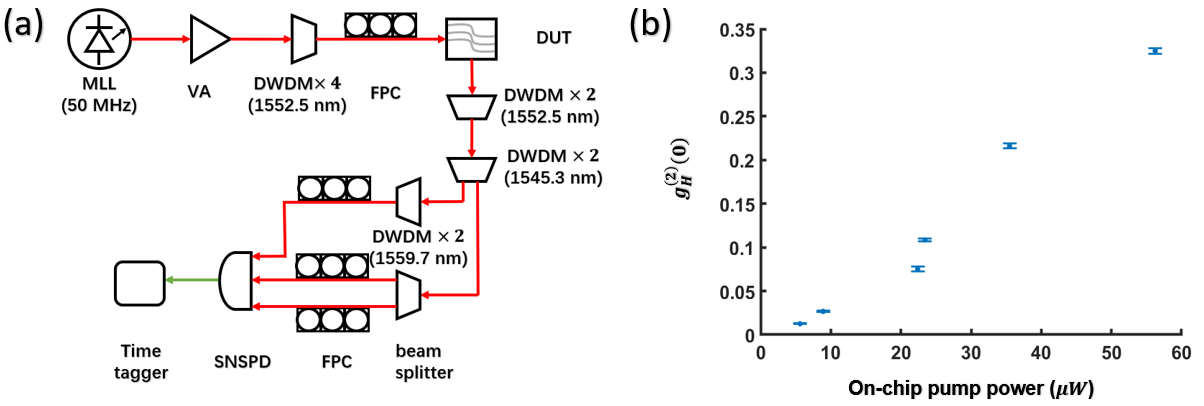}
\caption{\scriptsize{ (a) Experimental setup for heralded single-photon generation. (b) normalized auto-correlation function with zero time delay $g^{(2)}_H(0)$ at different on-chip pump powers}}
\label{figure4}
\end{figure}

The heralded single-photon generation and the auto-correlation function ($g_H^{(2)}(\tau)$) are important for ensuring the purity of photon pair source. As is shown in Figure \ref{figure4} (a), to measure the heralded single-photon generation, the idler is separated by a 50/50 splitter into two as idler I and idler II. Then together with signal, they are all sent into SNSPD to measure double coincidences between signal and Idler I ($N_{12}$), Idler II ($N_{13}$) and triple coincidences $N_{123}$. The auto-correlation function of $g_H^{(2)}(\tau)$ is then calculated by $N_{123}N_1/N_{12}N_{13}$, where $N_1$ donates signal detection rate. Using the programmable time-tagging unit, the coincidence rates at different time delays can be measured during the same experimental run. This is enabled by duplicating the detected physical photon channel to multiple virtual channels with different time delays and thus all corresponding coincidences are measured at once. This method largely simplifies the experiment and reduces noise and instabilities induced by different experimental runs. Figure \ref{figure4} (b) plots the auto-correlation function with zero time delay, which is normalized to the averaged $g^{(2)}_H(\tau)$ at other 20 ns-delays (corresponding to the 50 MHz repetition rate of our MLL). The lowest  $g^{(2)}_H(0)$ is measured to be 0.0126 $\pm$ 0.005 at an on-chip power of -22.5 dBm and $N_1$ of 18 kHz. It increases to 0.1085 $\pm$ 0.0014 at on-chip power of -16.3 dBm and $N_1$ of 310 kHz. This $g_H^{(2)}(\tau)$ measurement indicates our heterogeneous chip is a versatile single and paired photon sources for a variety of quantum optical information technologies.

Table \ref{table1} shows the state of the art of photon pair generation through $\chi^{(3)}$ process. Comparing with the previous result in a pure a-Si:H platform, our device has exhibited 4 times higher CAR and 30 times higher PGR due to much lower propagation/coupling loss through the waveguide. With the CAR higher than 1000, our work features a high PGR of 1.94 MHz. This shows that our device can massively produce high quality photon pairs. Leveraging high the nonlinearity of a-Si:H waveguide, the photon pairs can be produced efficiently through our non-resonant waveguide without the nonlinearity enhancement by microring. This allows more freedom of wavelength selection and avoid the instabilities of resonance shifting due to the thermo-optic effect or nonlinear phase modulation from a strong pump. Our showcase this heterogeneous a-Si:H/$Si_3N_4$ platform as an excellent and stable platform for photon pair and single photon generation.

\begin{table}[htbp]
 \centering \caption{The state of the art of photon pair generation through $\chi^{(3)}$ process}
\begin{tabular}{cccccc}
    \hline
   Reference& Material & On-chip pump power& CAR&PGR& $g_H^{(2)}$\\
   \hline
   K. Wang(2014)\cite{Wang:14}&a-Si:H &18 $\mu W$&399 &65 kHz&...\\
   This work& a-Si:H/$\mathrm{Si_3N_4}$& 0.9 $\mu W$&1632.6&1.94 MHz&...\\
   This work& a-Si:H/$\mathrm{Si_3N_4}$&5.6 $\mu$W&373.3&3.29 MHz&0.0126\\ 
   W. Jiang(2015)\cite{Jiang:15}&c-Si disk &79 $\mu W$&274&855 kHz&...\\
   X. Lu(2016)\cite{Lu:16}&c-Si disk&12$\mu W$&2610&1.2 kHz&0.003\\
  C. Ma(2017) \cite{Ma:17}&c-Si ring&7.4 $\mu W$&12105&16 kHz&0.005\\
  C. Ma(2017) \cite{Ma:17}&c-Si ring&59 $\mu W$&532&1.1 MHz&0.005\\
 X. Lu(2019) \cite{Lu:19}&$\mathrm{Si_3N_4}$ ring&46  $\mu W$&2280&4.8 kHz&...\\
  T. Steiner(2021)\cite{PRXQuantum.2.010337}&AlGaAsOI ring&3.4 $\mu W$&4389&230KHz&0.004\\

    \hline
   \end{tabular}
   \label{table1}
    \end{table}

\section{Conclusion}

In this paper, utilizing the high nonlinearity of a-Si:H for efficient photon pair generation and low loss of $\mathrm{SiN_x}$ for the routing of photon pairs, we have experimentally demonstrated photon-pair generation with a record high CAR of 1632.6 ($\pm$ 260.4) for non-resonant a-Si:H waveguides, which shows 4-times improvement compared to the previous single-layer a-Si:H platform \cite{Wang:14}. We also demonstrate multi-channel photon-pair sources with a CAR of around 200 across 10-nm bandwidth. Additionally, we have measured the heralded single-photon generation and demonstrated low correlation function of  $g_H^{(2)}(0)$ at 0.0125 $\pm$ 0.005. Our results show that with the increased coupling efficiency and low propagation losses of the $\mathrm{SiN_x}$ waveguide, higher photon detection rate can be observed with much lower on-chip pump power. This enables us to achieve a high CAR and high photon pair generation rate.  
The multi-material deposition and low-loss interlayer couplers has enabled a way to combine different materials' features onto the single platform and flexibly transfer photons and optical signals between different layers and materials with low transition loss. This solution offers  the possibility for the future PICs to integrate components from other material platforms and further miniaturize the experimental setups used in this work, for example $\chi^2$ nonlinearity and electro-optic modulation from lithium niobate, laser source from III-V materials, and photodetectors from germanium \cite{Chen:19,Wang2018,Jin:21,Keyvaninia:13,Guo_2019,Michel2010,MarrisMoriniVakarinRamirezLiuBallabioFrigerioMontesinosAlonsoRamosLeRouxSernaBenedikovicChrastinaVivienIsella+2018+1781+1793}.

\section{Acknowledgement}
This research was supported in part by National Science Foundation (\#1641094).

\bibliographystyle{plainnat}
\bibliography{main}

\end{document}